\documentstyle[12pt]{article}
\title{The effect of mixture lengths of vehicles on the traffic flow behaviour in one-dimensional cellular automaton} 
  
\author{H. Ez-Zahraouy$^*$, K. Jetto, A. Benyoussef\\\
\\
 Laboratoire de Magn\'{e}tisme et de la Physique
 des Hautes Energies
\\
Universit\'{e} Mohammed V, Facult\'{e} des Sciences, Avenue Ibn Batouta,  B.P. 1014
\\Rabat, Morocco
}
\date{ }

\begin{document}
\maketitle 
\abstract{The effect of mixture lengths of vehicles on the asymmetric exclusion model is studied using numerical simulations for both open and periodic boundaries in deterministic parallel dynamics. The vehicles are filed according to their length, the small cars type 1 occupy one cell whereas the big ones type 2 takes two. In the case of open boundaries two cases are presented. The first case corresponds to a chain with two entries where densities are calculated as a function of the injecting rates $\alpha1$ and $\alpha2$ of vehicles type 1 and type 2 respectively, and the phase diagram ($\alpha1$,$\alpha2$) is presented for a fixed value of the extracting  rate $\beta$. In this situation the first order transition from low to high density phases occurs at $\alpha1+\alpha2=\beta$ and disappears for $\alpha2>\beta$. The second case corresponds to a chain with one entry, where $\alpha$ is the injecting rate of vehicles independent of their nature. Type2 are injected with the conditional probability $\alpha\alpha2$, where $0\leq\alpha2=n\alpha\leq\alpha$ and n is the concentration of type2. Densities are calculated as a function of the injecting rates $\alpha$, and the phase diagrams ($\alpha$,$\beta$) are established for different values of n. In this situation the gap which is a characteristic of the first order transition vanishes with increasing $\alpha$ for n $\neq 0$. However, the first order transition between high and low densities exhibit an end point above which the global density undergoes a continuous passage. The end point coordinate depends strongly on the value of n. In the periodic boundaries case, the presence of vehicles type2 in the chain leads to a modification in the fundamental diagram (current, density). Indeed, the maximal current value decreases with increasing the concentration of vehicles type 2, and occurs at higher values of the global density.}\\
Pacs numbers: 05.50.+q , 64.60.Cn , 75.30.Kz , 82.20.wt 
\\$ ^*$corresponding author: ezahamid@fsr.ac.ma  
\newpage

\section{\protect\bigskip Introduction}

Without doubt, an efficient transportation system is essential for the functioning and success of modern industrialized societies. But the days when freeways were free are over. The increasing problems of roadway traffic raise the following questions: Is it still affordable and publicly acceptable to expand the infrastructure? Will drivers still buy cars when streets are effectively turned into parking lots? Automobile companies worried about their future market, have spent considerable amounts of money for research on traffic flow and on how the available infrastructure could be used more efficiently by new technologies.
Indeed, traffic flow is an interesting field of inter-disciplinary research; it has attracted the interest of many researchers from different disciplines. Like mathematicians, chemists, and engineers, physicists have addressed the problems of traffic flow many years ago and they have been trying to understand the fundamental principals governing the flow of vehicular traffic using theoretical approaches based on concepts and techniques of statistical physics\cite{1,2,3}. The approach of a physicist is usually quite different from that of a traffic engineer. A physicist tries to develop a model of the traffic by incorporating only the most essential ingredients which are absolutely necessary to describe the general features of typical real traffic. There are two different ways for modelling traffic: The macroscopic models which are based on fluid-dynamical description and the microscopic ones where attention is explicitly focused on individual vehicles which are represented by particles. The interaction is determined by the way the vehicles influence each others movement. In other words in the microscopic theories traffic flow is considered as a system of interacting particles driven far from equilibrium. Thus, it offers the possibility to study various fundamental aspects of the dynamics of truly non equilibrium systems which are of current interest in statistical physics\cite{4,5}. Within the conceptual framework of the microscopic approach, the particle hopping models describe traffic in terms of stochastic dynamics of individual vehicles which are usually formulated using the language of cellular automata (CA)\cite{6}. In general, CA are an idealization of physical systems in which both space and time are assumed to be discrete and each of the interacting units can have only a finite number of discrete states, thus in CA models of traffic the position, speed, acceleration as well as time are treated as discrete variables and the lane is represented by a one-dimensional lattice, each site represents a cell which can be either empty or occupied by at most one vehicle at a given instant of time. The computational efficiency of CA is the main advantage of this approach. Much effort has been concentrated on stochastic CA models of traffic flow first proposed by Nagel and Schreckenberg\cite{7}and subsequently studied by other authors using a variety of techniques\cite{8,9}. The stochastic dynamics of interacting particles have been studied in the mathematical and physical literature\cite{10}. In the physical case, driven lattice gases with hard core repulsion provide models for diffusion of particles through narrow pores and for hopping conductivity\cite{11}, and belong to the general class of non-equilibrium models which includes driven diffusing systems\cite{12,13}. They are closely linked to growth process\cite{14,15}, and can also be formulated as traffic jam or queuing problems\cite{16}. The fully asymmetric exclusion model (FAEM) which is also described in literature as the Totally asymmetric exclusion process (TASEP) corresponds to the case where particles hop only in one direction. This model can be divided into four classes according to the dynamics (sequential or parallel) and the choice of boundary conditions (open or periodic). In the sequential dynamics in which each particle has a probability $\Delta$t(time interval) of jumping to its right-hand neighbour if this neighbouring site is empty, has been solved exactly in one dimension with open boundary conditions\cite{17,18}. Recently Shaw et al \cite{19} studied the protein synthesis using the  TASEP with extended objects i.e. a system with particles of length $l>1$ using numerical and analytical tools for random sequential updating. The stochastic parallel update of the asymmetric exclusion model was studied using Monte-Carlo simulations\cite{20}, however  its deterministic case i.e. particles moving forward with probability $q=1$, was studied analytically by Tilstra and Ernst\cite{21}. The aim of this paper is to study the effect of mixture lengths $l=1$ and $l=2$ of vehicles on the FAEM in the case of deterministic parallel dynamics in both open and periodic boundary conditions using numerical simulations. This paper is organized as follows; section 2 is devoted to explain the model with three cases i.e. open boundaries with two entries, open boundaries with one entry and periodic boundaries. In the third section we present the main results obtained in these cases with a critical discussion. Section 4 is devoted to a conclusion.      

\section{Model}
We consider a chain of N sites in which the vehicles are filed according to their length. However, vehicles of type1 means short cars ($l=1$) which occupy one site and type2 the long vehicles ($l=2$) occupy two sites. Type2 moves by one site at each time step even if two sites ahead are empty, the same applies to type1. Hence Fig.1 displays an example of some configurations. 
\subsection{Open boundaries}
\subsubsection{The case of two entries}
In this case, type1 are injected in the first site (the first entry) with an injecting rate $\alpha1$, while type2 are injected in the third site (second entry) with an injecting rate $\alpha2$, if these entries are empty. However, in order to avoid the on-ramp at the moment when type2 are injected, we should take into account the following conditions:\\
The second site must be, also, empty, to not hinder the movement of type1 present in the first site at the next time step, because type2 enters the road on two time steps due to its length. Thus, the distance between the two entries should be at least equal to 1. 
\subsubsection{The case of one entry}
For both practical and theoretical reasons, some times different boundary conditions are used. Imagine a situation where a multilane road, containing different kinds of vehicles (car and trucks), is reduced to one lane due, for example, to a road construction. Such a situation can be modelled by using our model. In this case the multilane part of the road acts as a particle reservoir (both types), and the one part lane is represented by a chain with N sites. Here the vehicles will be injected at the same entry depending on the injecting rate $\alpha$ of vehicles and the concentration n of vehicles type2. In this case, numerical simulations are limited only when $0\leq n \leq 0.5$, since in the road we have a few type2 vehicles in comparison to type1 (we assume that type2 are trucks and type1 are cars). Hence, we denote by $\alpha2$= n$\alpha$, the injecting rate of vehicles type2. However, the process of the injection of the vehicles is as follows:\\
If the first site is empty, a random number $0\leq R\leq 1$ is chosen. Then, if $R\leq\alpha$, a vehicle is injected. But its nature depends on the value of $\alpha2$; hence, if $\alpha2\leq\ R \leq\alpha$, then type1 could enter the road; while, if $ R \leq\alpha2$, type2 can enter the road.\\
Note that $0\leq\alpha2\leq\alpha$, in contrast with the first boundary case where $0 \leq\alpha2 \leq 1$.
 
\subsection{periodic boundaries}
The periodic boundary conditions for the FAEM with one species of particle on a ring has been studied exactly by Schadschneider and Schreckenberg\cite{8,9}. Here we study the FAEM with two types of vehicles for periodic boundaries with deterministic parallel dynamics. To illustrate this situation let us consider a ring of N sites where $C_{1}$ and $C_{2}$ are  the densities of sites occupied by the vehicles type1 and type2 respectively. Hence, the global density is given by $C=C_{1}+C_{2}$. We assume that $n=\frac{\rm C_{2}}{\rm C}$ is the concentration of the sites occupied by type2. However, $C_{1}=(1-n)C$ and $C_{2}=nC$. The main idea of studying this case is to establish the fundamental diagram (I,C) for several values of n. The mean value of the current in the steady state is given by: 
$$I=\frac{\rm 1}{\rm NT}\sum_{t=1}^{T}\sum_{i=1}^{N}I_{i}$$

\section{Simulations and results}
In our computational studies we have considered a chain with $N=1000$ sites. Simulation of the closed system begins with particles randomly distributed around the ring depending on their densities $C_{1}$ and $C_{2}$, whereas the open system begins with a small number of vehicles randomly distributed in the chain. The systems run for 20 000 MCS to ensure that steady state is reached for the periodic case and 40 000 MCS for open systems, at this moment data including the current and density are collected. In order to eliminate the fluctuations 25 initial configurations were randomly chosen.  
\subsection{Open boundaries} 
\subsubsection{The case of two entries}
We recall that our aim in this case is to explore the phase diagram ($\alpha1$,$\alpha2$) for a fixed values of the extracting rate $\beta$.
Fig.2 shows the profile of the global density $\rho$ (which is the density of all sites occupied by both types of vehicles) versus $\alpha1$ for $\beta=0.3$ and various injecting rates $\alpha2$. We distinguish two regions:\\
The first region is when $(\alpha1+\alpha2)<\beta$ in which $\rho$ increases with $\alpha1$, but this increase depends on the value of $\alpha2$. Indeed: 
$$\rho(\alpha1,\alpha2=0) < \rho(\alpha1,\alpha2=0.1)< \rho(\alpha1,\alpha2=0.2)$$
On the other hand this region corresponds to the low density phase due to the low values of $\rho$.  
The second region occurs when $(\alpha1+\alpha2)>\beta$ in which $\rho$ become constant (about 0.77) and does not depend on $\alpha1$ and $\alpha2$. This is the high density phase.\\
At $(\alpha1+\alpha2)_{c}=\beta$ the global density $\rho$ is discontinuous and undergoes a jam between the two regions which is a characteristic of the first order transition.\\
We can see that the system undergoes the usual first order transition at $\alpha1=\beta$ for $\alpha2=0.0$ in agreement with\cite{22}. Also we remark that the density takes its higher values when $\alpha2 > \beta$ even for $\alpha1=0.0$, this means that there is no low density phase in this case i.e. the system prefers to be in the high density phase. In fact type1 and type2 are not correlated at the beginning of the road due to the conditions mentioned in section (2.1.1), this explains why the first order transition occurs at$(\alpha1+\alpha2)_{c}=\beta$. For $\alpha2 > \beta$ this first order transition disappears due to the greater proportion of type2 in the road, indeed the number of type2 which enter the chain is greater than those which leave it leading to high density phase independently of $\alpha1$.  
Collecting all of these results we obtain the phase diagram $(\alpha1,\alpha2)$ for $\beta=0.3$ in Fig.3 The same results can be obtained for others values of $\beta$.        
\subsubsection{The case of one entry}
This situation corresponds to the case when type1 and type2 enter the chain in the same entry. We recall that the FAEM with open boundary conditions and deterministic parallel dynamics exhibits a first order transition \cite{22}, in which $\rho$ is discontinuous and undergoes a gap between the low and high density phases at $\alpha_{c}=\beta$. In our model the gap decreases with increasing $\alpha$ and vanishes for its higher values. Fig.4 shows the profile of the global density $\rho$ versus $\alpha$ for $\alpha2=n\alpha$ and various extracting rates $\beta$, here n=1/4 corresponding to the concentration of type2. We remark that for $\beta=0.0-0.3$ the first order transition occurs at $\alpha_{c}=\beta$ in agreement with \cite{22} and at $\alpha_{c}<\beta$ when $\beta=0.4-0.7$. 
For $\beta=0.8-0.9 $, there is no phase transition. In fact, in this case about 1/4 of the vehicles which enter the road could be type2, and their number becomes important as long as $\alpha$ increases, we recall that types2 occupy two sites due to their length so, it is clear in this situation that the global density in the road becomes larger than the n=0 case (i.e. the chain contains only type1 vehicles). Such a comparison is illustrated in Fig.5, in which the n=1/4 density gap is smaller than the n=0 one. The first order transition occurs at$\alpha_{c}<\beta$(n=1/4) instead of $\alpha_{c}=\beta$(n=0.0). This result is not observed when $\beta=0.1- 0.3$(Fig.4), because the system reach the high density phase for small values of $\alpha$, which means that the probability for type2 to enter the road is small. Furthermore, for $\beta=0.8-0.9$, the gap vanishes (Fig.4) and we have a continuous passage from the low to the high density phase. In Fig.6 we display the $(\alpha,\beta)$ phase diagram for three values of n. It is found that the first order line transitions, separating the low and high density phases in the case of $n=0.0$ \cite{22}, is modified when $n=0.25,0.5$. Indeed, for $n=0.25$ and $n=0.5$ the line transitions terminate by end points above which the first order transitions disappear. These end points are located at $(\alpha=0.7,\beta=0.74)$ and $(\alpha=0.5,\beta=0.57)$ for $n=0.25$ and $n=0.5$ respectively. 
\subsection{Periodic boundaries}
The periodic boundary conditions for the FAEM with one species of particle (Type1) on a ring has been studied exactly by Schadschneider and Schreckenberg\cite{8,9}, where the relation between current and density is given as follows: $$I(C,P)=1/2[1-\sqrt{1-4qC(1-C)}]$$
p is the braking probability and $q=1-p$ is the hopping rate.\\
An interesting feature of this expression is that the current is invariant under the operation $C \longrightarrow (1-C)$ which interchanges particles and holes. Therefore the fundamental diagram is symmetric about $C_{m}=1/2$. This symmetry is conserved for all p, but breaks down for a fixed value of p and $V_{max}>1$ where the magnitude of $C_{m}$ decreases and the current increases with increasing $V_{max}$.\\
In our model, introducing a small concentration of type2 leads to a breaking  of the symmetry between the holes and particles and the expression of the current $I(C,P)$ mentioned above doesn't work. Fig.7 shows the fundamental diagram (I,C) for different concentrations n of type2, it is clear that the magnitude of $C_{m}$ increases and the current decreases when n increases. We can say that so long as C is sufficiently small the vehicles are too far apart to interact mutually, therefore the current increases with C but this increase is controlled by the concentration n. Indeed type2 takes lot of time in passing the local detector due to its length in contrast to type1, which reduces the current and shifts to the right $C_{m}$. In the particular case $n=1$, with a random sequential dynamic, Shaw et al\cite{19} found using analytical and numerical calculations, that the current I isn't symmetric about $C_{m}=1/2$. Indeed, the maximal current is lowered from $0.25$(n=0, the chain contains only type1 vehicles) to $I_{m}=0.1715$(n=1) and the density $C_{m}$ is shifted from $0.5$ to $C_{m}=0.585$. However, using numerical simulations with a deterministic parallel dynamics update, we find that the maximal current is lowered from $0.5$(n=0)\cite{8,9} to $I_{m}=0.332$ (n=1) and the density $C_{m}$ is shifted from $0.5$ to $C_{m}=0.669$ as presented in Fig.7.
\\ We note that Fig. 7 can be reproduced analytically from the expression of the current cited above (for q=1) by rescaling the density C and the current I. Indeed, instead of C one then has $C(1-n/2)/(1-nC/2)$ and for the current I one gets $I/(1-nC/2)$, hence the expression of the current becomes $ I(C)=1/2[1-nC/2-\mid 2C-1-nC/2 \mid]$. 

\section{Conclusion}
We have studied the effect of a mixture of lengths of vehicles on the traffic flow using numerical simulation with a deterministic parallel update. For this purpose we have defined two types of vehicles depending on their length, type1 the small cars ($l=1$) which occupy one site, and type2 the long ones ($l=2$) taking two sites. We have investigated both open and periodic boundaries. In the former case, two different cases are considered. In the case of two entries, the global density $\rho$ is discontinuous and undergoes a jam between the low and high density phases which is a characteristic of the first order transition at ($\alpha{1}+\alpha{2})_{c}=\beta$; and for $\alpha2 > \beta$ the system is in the high density phase then  no phase transitions is observed. In the case of one entry, the concentration of type2 plays a major role, indeed the gap between low and high density phases decreases when increasing $\alpha$ and vanishes for higher values. Furthermore the phase diagram exhibits an end point above which the first order transitions disappear.\\
In the periodic boundary case, the presence of type2 vehicles leads to a breaking of the symmetry holes-particles, then the magnitude of $c_{m}$ increases, shifting to high density and the maximal current decreases.
    
\par\bf  ACKNOWLEDGMENT \\
\par \rm  This work was financially supported by the Protars II n° P11/02\\

\newpage \textit{Figure captions:}

Fig. 1: Example of configurations obtained after three steps for system size $L=8$ where both type1 and type2 move with $V_{max}=1$.

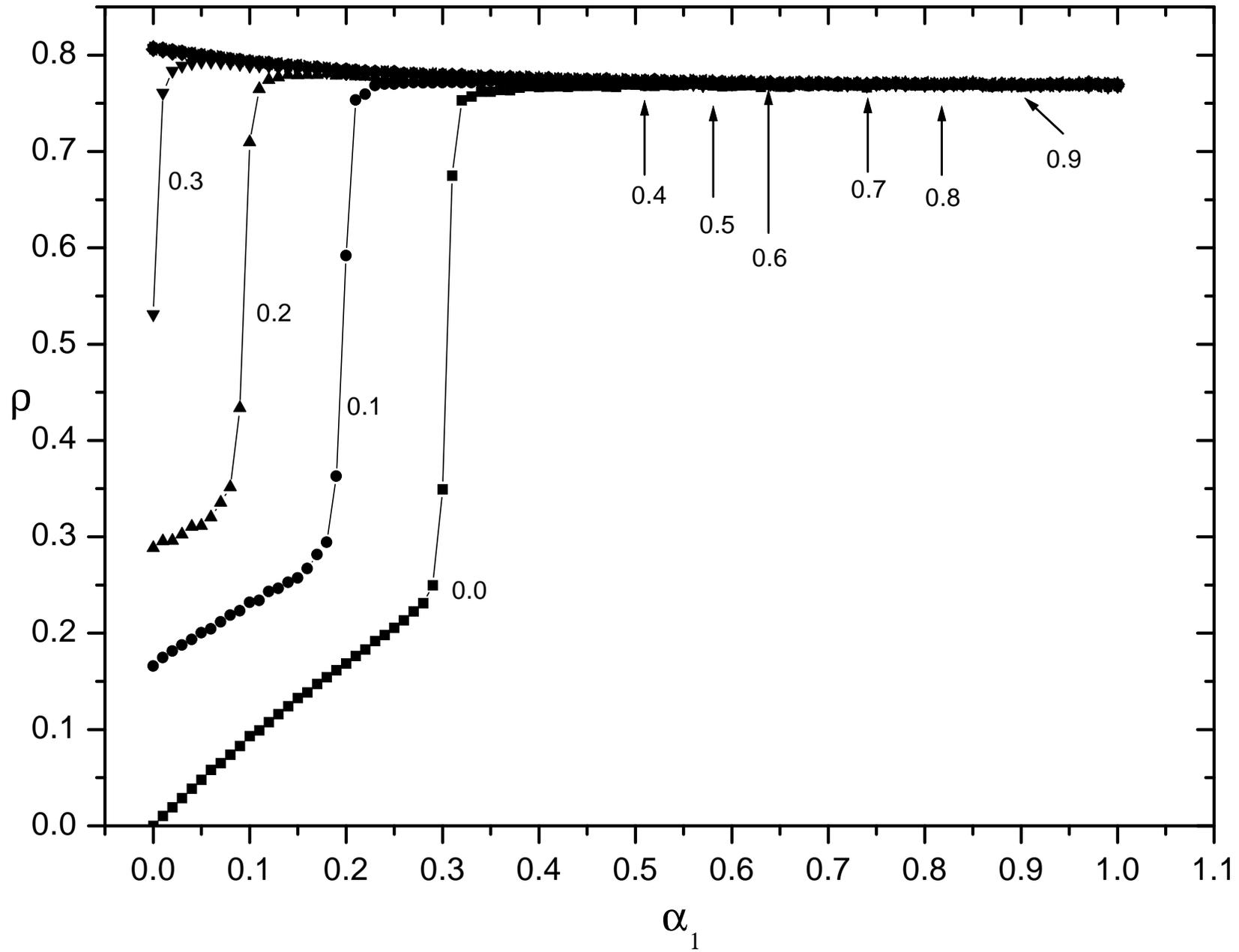
Fig. 2: The variation of the global density $\rho $ versus $\alpha1$ for different values of $\alpha2$ and $\beta=0.3$ in the case of open boundaries with two entries, the number accompanying each curve denotes the values of $\alpha2$.

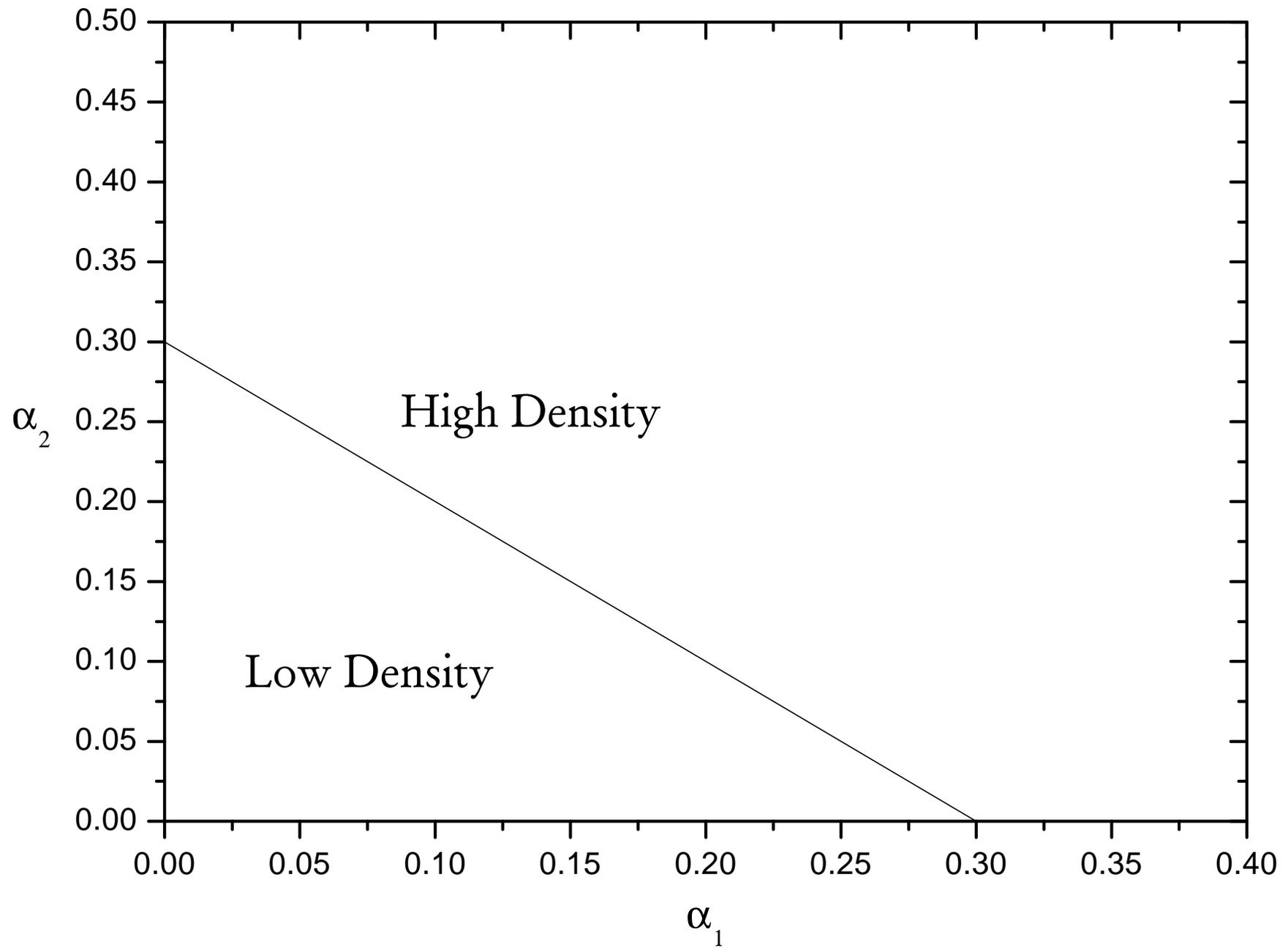
Fig. 3: The phase diagram $(\alpha1,\alpha2)$ for $\beta=0.3$ in the case of open boundaries with two entries.

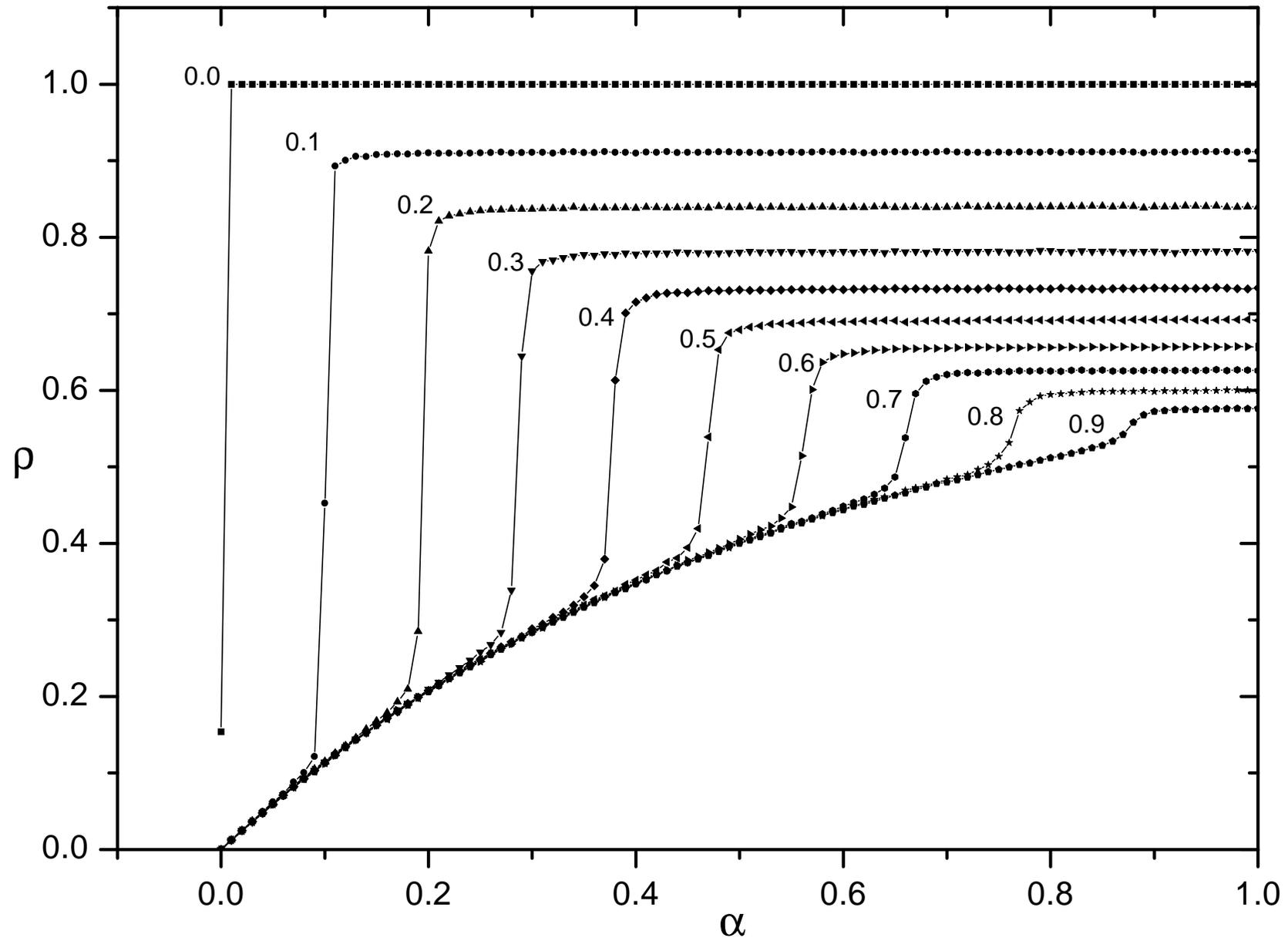
Fig. 4: The variation of the global density $\rho $ as a function of $\alpha $ with $\alpha2=1/4 \alpha$ and for different values of $\beta$ in the case of open boundaries with one entry. The number accompanying each curve denotes the value of $\beta$.

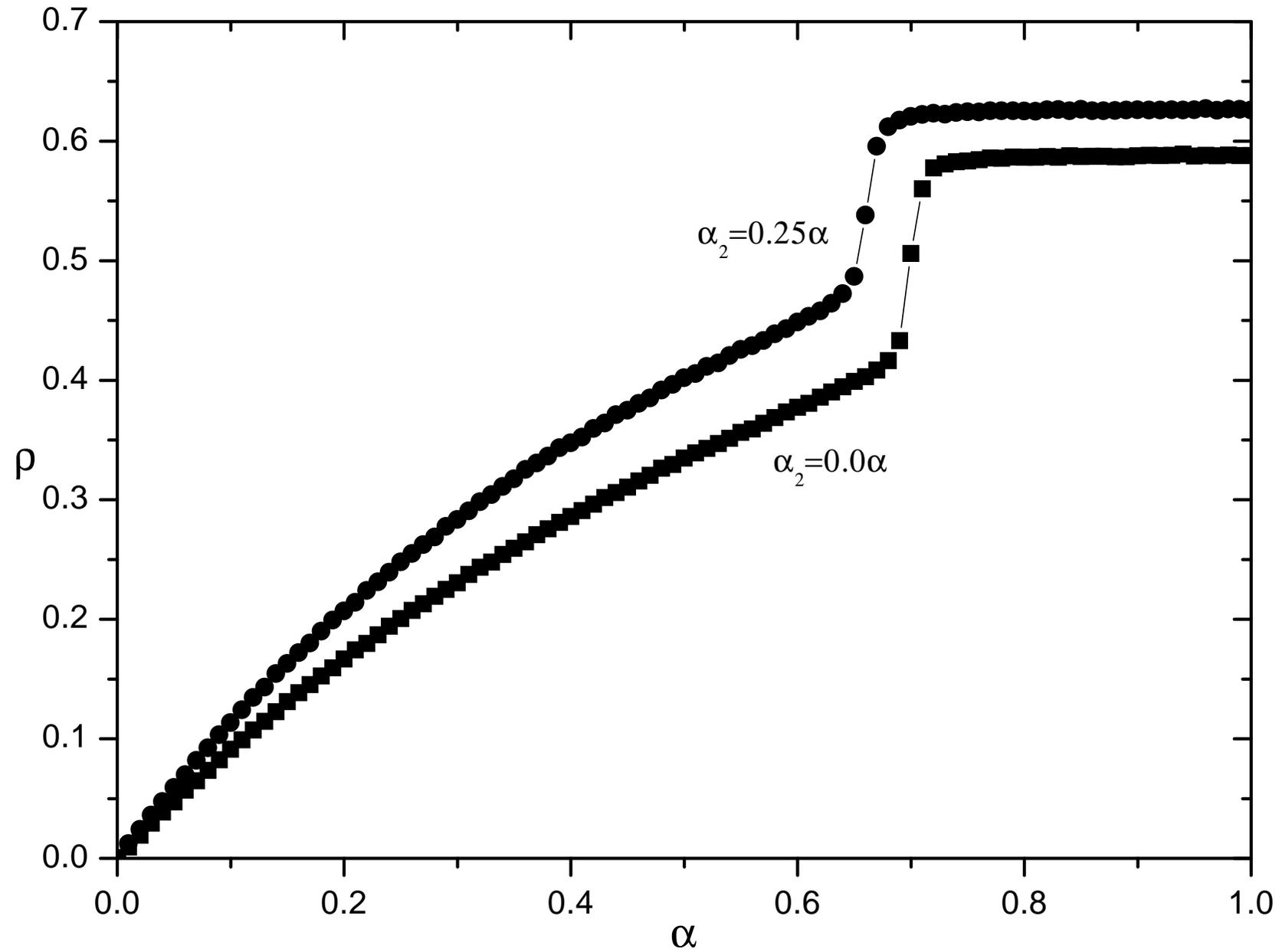
Fig. 5: The variation of the global density $\rho $ as a function of $\alpha $ with $\alpha2=1/4 \alpha$(circle dot) and $\alpha2=0.0\alpha$(square dot) for $\beta=0.7$ in the case of open boundaries with one entry.

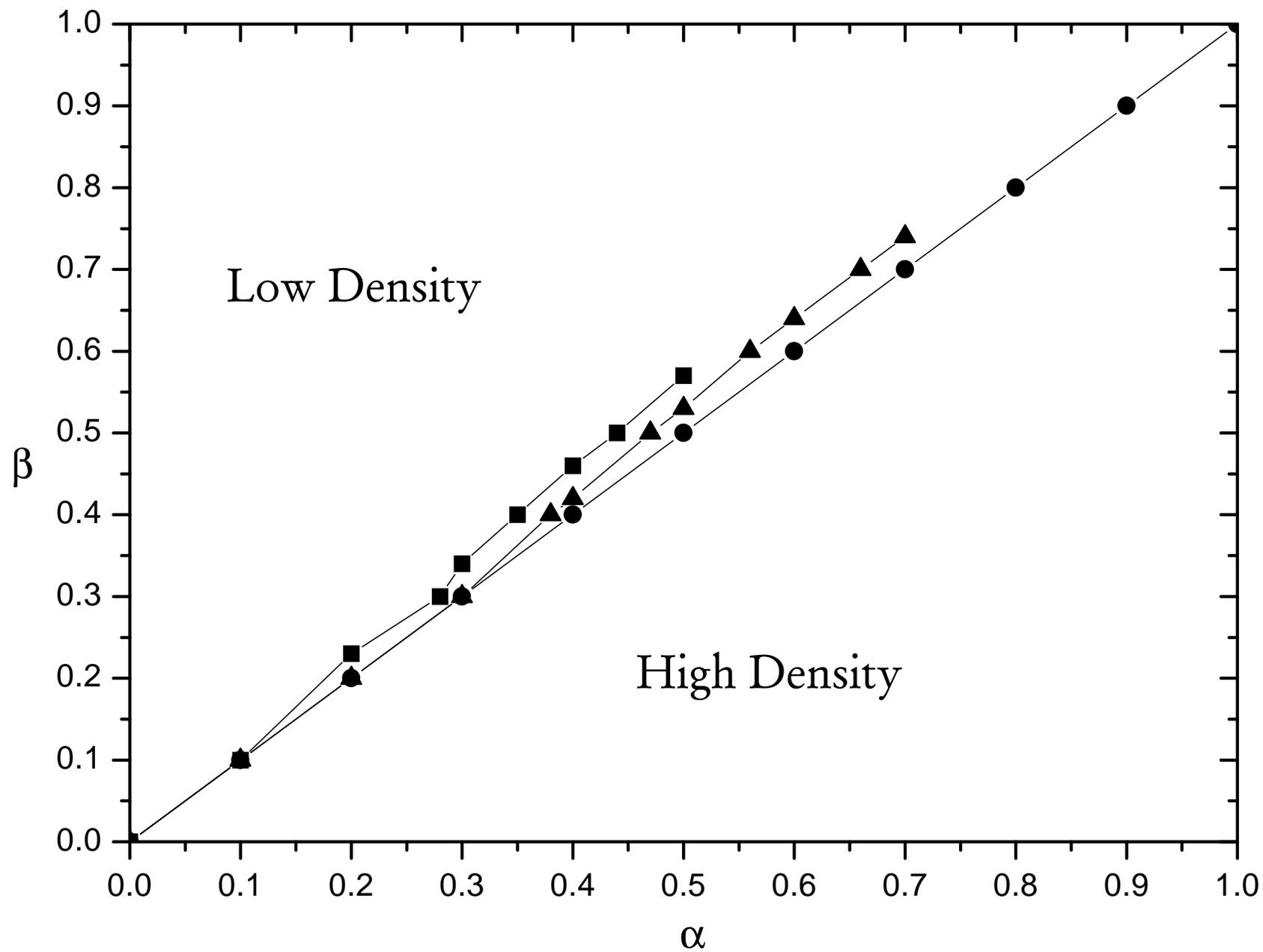
Fig. 6: Phase diagram in the $(\alpha ,\beta)$ plane for several values of n(concentration of type2). Square and up triangle dots correspond respectively to $\alpha2=0.5\alpha$ and $\alpha2=0.25\alpha$ where the lines of the first order transitions ended at the end points (0.5,0.57) and (0.7,0.74). Circle dots recall the case $\alpha2=0.0\alpha$.    

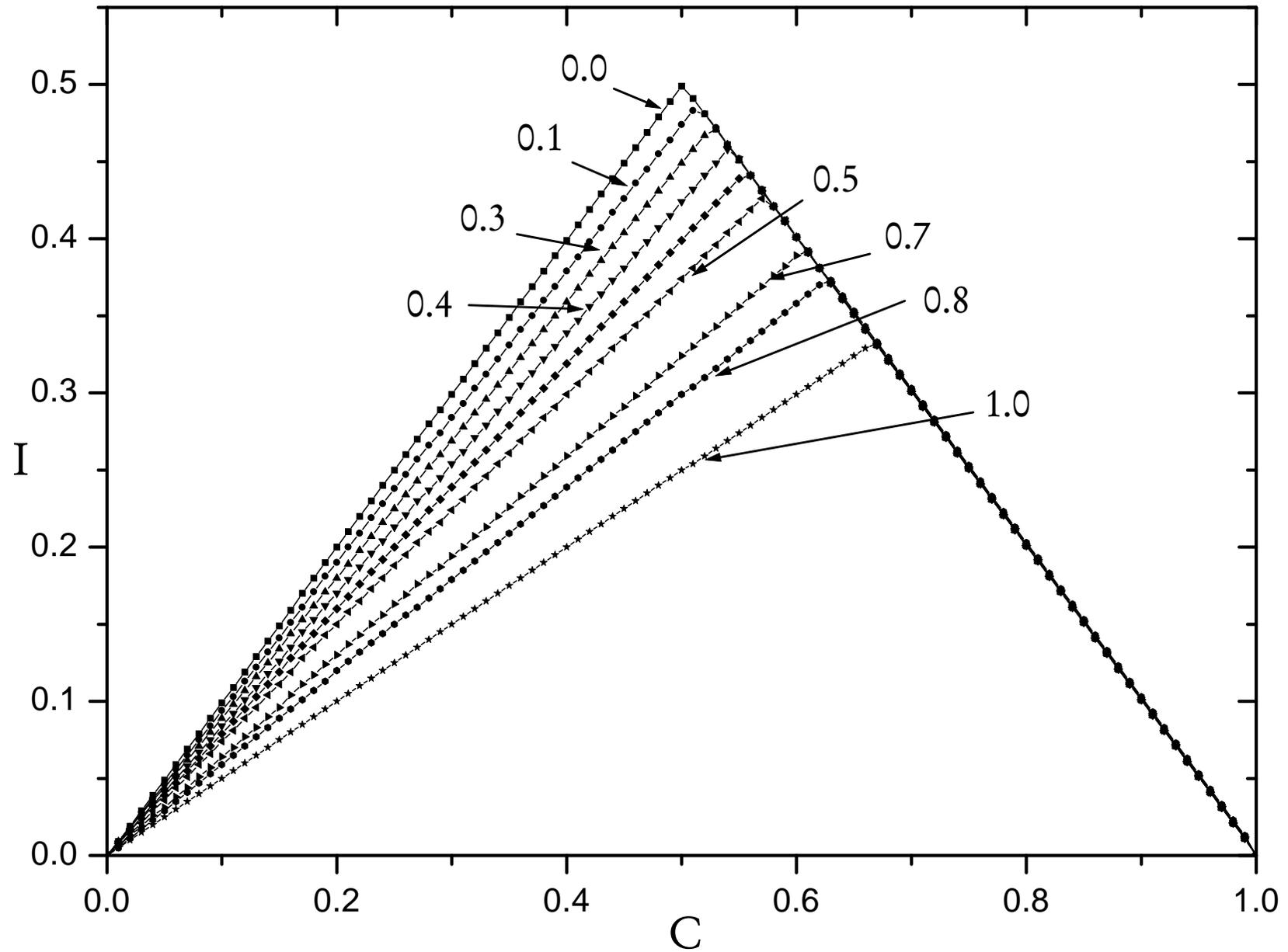
Fig. 7: The fundamental  diagram (I,C) for several values of n. The symbols are obtained by simulation whereas the lines represent the analytical expression of the current.

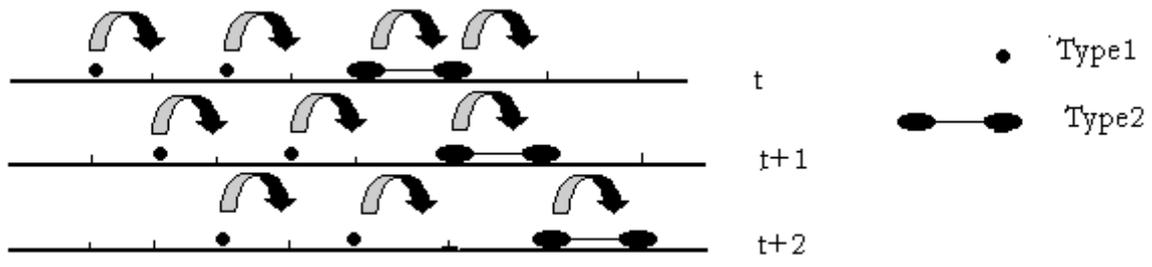
\end{document}